\documentclass[twocolumn,showpacs,amssymb,prd]{revtex4}

\usepackage{graphicx}
\usepackage{dcolumn}
\usepackage{bm}

\def\lapproxeq{\lower .7ex\hbox{$\;\stackrel{\textstyle
<}{\sim}\;$}}
\def\gapproxeq{\lower .7ex\hbox{$\;\stackrel{\textstyle
>}{\sim}\;$}}

\begin{document}

\hyphenation{Es-ta-du-al}

\preprint{??-??-????}

\title{High energy neutrinos from radio-quiet AGNs} 

\author{Jaime Alvarez-Mu\~niz$^{1,2}$ and Peter M\'esz\'aros$^{1,3,4}$} 

\affiliation{
$^1$Dept. de F\'\i sica de Part\'\i culas, 
Universidade de Santiago de Compostela,
15706 Santiago de Compostela, A Coru\~na, Spain \\ 
$^2$Dept. of Astronomy \& Astrophysics,
Pennsylvania State University, University Park, PA 16802, USA \\ 
$^3$ Dept. of Physics, 
Pennsylvania State University, University Park, PA 16802, USA \\
$^4$ Institute for Advanced Study, Princeton, NJ 08540, USA}


\begin{abstract}
Most active galactic nuclei (AGN) lack prominent jets,
and show modest radio emission and significant X-ray emission
which arises mainly from the galactic core, very near from the
central black hole. We use a quantitative scenario of such 
core-dominated radio-quiet AGN, which  attributes a substantial
fraction of the X-ray emission to the presence of abortive jets
involving the collision of gas blobs in the core. Here we investigate 
the consequences of the acceleration of protons in the shocks from
such collisions. We find that protons will be accelerated up to 
energies above the pion photoproduction threshold on both the X-rays 
and the UV photons from the accretion disk. The secondary charged 
pions decay, producing neutrinos. We predict significant fluxes of 
TeV-PeV neutrinos, and show that the AMANDA II detector is already 
constraining several important astrophysical parameters of these 
sources. Larger cubic kilometer detectors such as IceCube will be 
able to detect such neutrinos in less than one year of operation, 
or otherwise rule out this scenario.
\end{abstract}


\maketitle


\section{\label{introduction}Introduction}

Around 90\% of all active galactic nuclei (AGN), such as
quasars and Seyfert galaxies, are radio-quiet, i.e.  their 
radio emission is relatively modest. This is due to the 
lack of prominent jets which provide most of the radio 
emission in the remaining 10\% of AGNs classified as radio-loud. 
Both types of AGN are also strong X-ray emittors, and the ultimate 
energy source is thought to derive from accretion of gas onto a 
central massive black hole (e.g. see \cite{poutanen99}). The high 
energy photon emission in radio-loud jet-dominated sources such 
as Blazars is largely due to non-thermal processes in the jet, 
whose dimensions extend well beyond the outer edge of the galaxy. 
In radio-quiet AGNs, on the other hand, the short X-ray variability 
timescales as well as spectral analysis of Fe X-ray lines indicate 
that the emission arises in the galactic core, very near from the 
central black hole. Both types of AGN have been considered as 
possible sources of high energy cosmic rays and neutrinos (see
e.g. \cite{hooper02} and refs. therein). 
Here we concentrate on the more numerous radio-quiet AGN as a 
source of high energy neutrinos.

The typical model of radio-quiet AGN X-ray emission is based on
UV photons from the accretion disk suffering repeated inverse 
Compton scatterings in a hotter corona, leading to a power law 
X-ray spectrum (e.g. \cite{zdziarski00}). This generic model 
does not allow for recent evidence indicating that, at some level, 
many radio-quiet AGN show weak radio emission, and in rare cases 
even show a weak incipient radio jet. A quantitative model which 
attempts to provide a bridge between the two classes of AGN 
considers the high energy emission in the core 
of radio-quiet AGNs due to the presence of an abortive jet
\cite{Ghisellini04}. These could arise from an 
outflow from the innner edge of the accretion disk, similar to that 
inferred in radio-loud objects, but  whose velocity is smaller than 
the escape velocity from the central black hole region. As in 
succesful jets, the outflow is expected to be intermittent, leading 
to shells or blobs ejected from the central region. In this case, 
due to their lower velocity, they will reach a maximum distance 
from the hole before falling back, occasionally colliding with 
subsequently ejected, outgoing blobs. The X-ray emission in this 
model is attributed to Inverse Compton scattering of UV photons 
from the accretion disk by electrons accelerated in the 
sub-relativistic shocks resulting from blob-blob collisions. 

Here we consider the consequences of the acceleration of protons
in the same shocks which accelerate the electrons. We find that 
protons will be accelerated 
up to energies above the energy threshold for pion photoproduction 
on both the X-rays produced by the accelerated electrons, as well 
as on the UV photon field from the accretion disk. The charged pions 
decay producing neutrinos. A natural outcome of this model is then 
the production of a flux of high energy (TeV-PeV) neutrinos, which
may be detectable with $\sim {\rm km^3}$ volume detectors such as 
Icecube \cite{icecube}. Radio-quiet AGN constitute the main bulk 
of AGNs, and on average can be found closer to Earth. Also, since
the neutrinos are produced in sub-relativistic shocks the emission 
is isotropic and the aborted jets need not point towards the Earth 
in order for the neutrinos to be detectable. We stress, however,
that our conclusions are not restricted to this specific model
of aborted jets, although it is used here as a framework. 
Qualitatively similar results might in principle be expected in 
disk-corona or other AGN models, if magnetic reconnection (e.g.
\cite{nayakshin00}) or shocks associated with the accretion
process (e.g. \cite{kazanas86},\cite{stecker96}) lead to proton 
acceleration near the central black hole.

In this paper we first show in Sec. \ref{acceleration} that protons 
can be accelerated up to $E_p^{max}\approx 10^9$ GeV in the 
blob-blob collision shocks. We discuss proton interactions with 
X-rays, UV photons and cold unaccelerated protons in Sec. 
\ref{interactions}. In Sec. \ref{neutrinos} we discuss neutrino 
production by photomeson and $pp$ interactions, and we estimate 
the neutrino flux from a typical radio-quiet quasar as well as 
the cumulative diffuse neutrino flux from all the radio-quiet
quasars in the Universe. We also 
calculate in this section the muon and shower event rate expected in 
Gigaton Cherenkov detectors such as Icecube. In Sec.\ref{conclusions} 
we summarize the paper and discuss the implications of our findings.

\section{\label{acceleration}Proton acceleration and cooling times}

\begin{figure}
\centerline{\includegraphics[width=8.5cm]{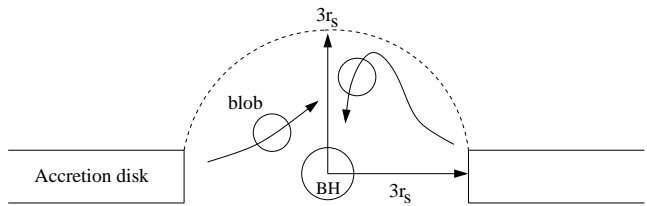}}
\caption{Sketch of the geometry of a radio-quiet AGN core 
adopted in this work, showing the $10^8\;M_\odot$ black hole (BH), 
accretion disk and two blobs moving
in opposite directions which are about to collide.
$r_s\approx 2.95\times 10^{13}$ cm is
a Schwarzschild radius for a $10^8 M_\odot$ black hole.}
\label{fig:agn_cartoon}
\end{figure}

We adopt here a nominal set of parameters which serve as a specifc 
example for the purposes of normalization. We consider an AGN with 
a $10^8$ solar mass black hole and with an accretion disk inner edge 
at a distance of about $3r_s$ from the black hole, where $r_s=2GM/c^2 
\approx 2.95 \times 10^{13}$ cm is the Schwarzschild radius
(Fig. \ref{fig:agn_cartoon}). 
We assume that blobs are ejected with an initial velocity 
$\beta_0=0.5$, which corresponds to a Lorentz factor $\Gamma_0\sim 1.15$,
i.e. they are non-relativistic and we assume 
that $\Gamma_0\sim 1$ from now on, and hence we do not make any
distinction between the comoving (blob) frame and the 
observer's frame. The size of a blob is $r_{\rm blob}\approx\alpha_r r_s$
with $\alpha_r\sim 1$. The lifetime of the shock produced in a 
blob-blob collision is roughly equal to the blob-blob crossing time, 
$t_{\rm cross}\approx \alpha_r r_s/c\beta_0\approx 2.0\times 10^3\;{\rm s}$. 

The Eddington luminosity of a $10^8 M_\odot$ black hole AGN is 
$L_{Edd}=4\pi GM m_p c/\sigma_T \approx 1.26 \times 10^{46}\;L_{Edd,8}
\;{\rm erg\;s^{-1}}$. This corresponds to an Eddington accretion rate 
$\dot M_{Edd}=L_{Edd}/\eta c^2\approx 2.8 M_\odot L_{Edd,8}
\;\eta_{0.08}^{-1}\;{\rm yr^{-1}}$, where $\eta=0.08\;\eta_{0.08}$.
Following \cite{Ghisellini04} we assume that the accretion
results in the intermittent ejection of blobs of matter. We also 
assume that a significant fraction $\alpha_{bb}\sim 1$ of the observed 
X-ray luminosity is due to the blob-blob collisions, whose number at 
any given time is $N_{\rm bb}$. $N_{\rm bb}$ can be estimated through a simple 
geometrical argument, from the number of blobs ($N_{\rm blob}$)
that are present in the region around the BH at that given time. 
$N_{\rm blob}$ is given roughly by the ratio of the time of flight 
of a blob i.e. the time it takes for a blob to go upwards from the 
BH and to fall back onto it, $t_{\rm flight} \approx 6r_s/\beta_0 c$, 
to the average time between the launch of consecutive blobs $\Delta t$. 
Taking $\Delta t\approx (1-2)\;r_s/c$, a value that qualitatively 
accounts for the time variability of radio-quiet quasars 
\cite{Ghisellini04}, we obtain $N_{\rm blobs}\approx 10$. 
The average number of collisions (optical depth) of a blob during
its flying time is given by the 
ratio of the blob flying time $t_{\rm flight}\sim V_T^{1/3}/(\beta_0 c)$ 
to the mean time between blob collisions, 
$V_T /(N_{blob} V_{blob}^{2/3} \beta_0 c)$, where the volume of a blob 
is $V_{\rm blob}\approx (4/3)\pi r_s^3$ and the total volume 
inhabited by the blobs  is $V_T\approx (4/3)\pi (3r_s)^3$
(see Fig. \ref{fig:agn_cartoon}), giving an average number of 
collisions per blob $\approx 1$. The number of blob-blob 
collisions at any given time is then $1<N_{bb}<5$. 
We then assume that the X-ray luminosity from the steady-state
number $N_{\rm bb}$ of blob-blob collisions is equal to a fraction 
$\alpha_{\rm bb}\sim 1$ of the total X-ray luminosity $L_{\rm X}$. 
Assuming $L_{\rm X}$ to be a fraction $\alpha_{X,Ed}<1$ of the 
Eddington luminosity, we combine these two factors into a single 
efficiency $\alpha_{\rm X} =L_{\rm X} / L_{Edd}$ which we parametrize 
as $\alpha_{\rm X}\approx 0.01 \;\alpha_{0.01}$ corresponding to  
$L_{\rm X}\approx 10^{44}\;{\rm erg\;s^{-1}}$, which is the 
X-ray luminosity of a nominal radio-quiet AGN \cite{Boyle93}. 
All the calculations performed are parametrized by  
$\alpha_{\rm X}$, which can vary in order to adjust $L_{\rm X}$ 
to the actual observed X-ray luminosity from a particular AGN.

Using the usual assumption that the magnetic field $B$ in the shock 
is due to turbulent fields generated behind the shocks arising
from the blob-blob collisions, the comoving (in the blob frame) 
magnetic field energy density is a fraction $\epsilon_B$
of the post-shock proton thermal energy:
\begin{equation}
\frac{B^2}{8 \pi} \approx \epsilon_B \rho_{\rm preshock} c^2 
\end{equation}
The density $\rho_{\rm preshock}$ should be roughly equal to the 
density of the accretion disk at a radius $\approx 3r_s$, which 
is given by \cite{Shakura73},
\begin{eqnarray}
\rho_{\rm preshock} &\approx &
\rho_{\rm disk}(3r_s)=\frac{\dot M}{4\pi(3r_s)^2 \alpha_v \sqrt{GM/3r_s} } 
\nonumber \\
&\approx & 1.5 \times 10^{-14}\; \alpha_{\rm X,0.01} 
L_{Edd,8}\;\eta_{0.08}^{-1}  
\;\; {\rm g\; cm^{-3}}
\end{eqnarray}
where $\dot M=\alpha_{\rm X} L_{Edd}/\eta c^2$ and 
we have taken the viscosity $\alpha_v\approx 0.1$ \cite{Shakura73}.
From $\rho_{\rm preshock}$ the proton number density is,
\begin{equation}
n_p=\frac{\rho_{\rm preshock}}{m_p}\approx 8.7 \times 10^{9}\;
\alpha_{\rm X,0.01} L_{Edd,8}\; \eta_{0.08}^{-1} \;\;\; {\rm cm^{-3}}
\label{eq:pdensity}
\end{equation} 
where $m_p$ is the proton mass. The magnetic field is then 
\begin{eqnarray}
B &\approx & \sqrt{8\pi}\epsilon_B^{1/2}
\rho_{\rm preshock}^{1/2} c \nonumber \\
&\approx & 1.0 \times 10^4 \; (\alpha_{\rm X,0.01} L_{Edd,8})^{1/2}
\epsilon_{B,0.3}^{1/2} \;\;\; {\rm G}
\label{eq:Bfield}
\end{eqnarray}
where $\epsilon_B=(1/3)\;\epsilon_{B,0.3}$.

The proton acceleration time is given by $t^{acc}_p\approx A_p r_L/c$ 
where $r_L$ is the Larmor radius in the magnetic field in the blob,
and $A_p$ is a constant $\geq 1$. Using Eq. (\ref{eq:Bfield}) we get,
\begin{eqnarray}
t^{acc}_p &\approx & \frac{A_p \gamma_p m_p}{eB} \nonumber \\
&\approx & 9.9\times 10^{-9} \; A_p 
(\alpha_{\rm X,0.01} L_{edd,8})^{-1/2} 
\epsilon_{B,0.3}^{-1/2} \; \gamma_p \;\;  {\rm s}
\label{eq:acceleration}
\end{eqnarray}  
The proton synchrotron cooling time is 
\begin{eqnarray}
t^{syn}_p &=& \frac{6\pi m_p^3 c}{\sigma_{\rm Th} 
m_e^2 \gamma_p B^2} \nonumber \\
&\approx & 4.4 \times 10^{10} \;  
(\alpha_{\rm X,0.01} L_{Edd,8})^{-1} \epsilon_{B,0.3}^{-1} \; \gamma_p^{-1}
\;\;\; {\rm s}
\label{eq:synchrotron}
\end{eqnarray}
where $\sigma_{\rm Th}\approx 0.665 \times 10^{-24}\;{\rm cm^2}$ 
is the Thomson cross section. Equating the synchrotron cooling time
and the acceleration time (see Eq. (\ref{eq:acceleration})) we get the 
maximum energy of the accelerated protons,
\begin{equation}
E^{max}_p \approx 2.0 \times 10^9 \; A_p^{-1/2}  
(\alpha_{\rm X,0.01} L_{Edd,8})^{-1/4} \epsilon_{B,0.3}^{-1/4} \;\;\; {\rm GeV}
\label{eq:Epmax}
\end{equation}

Protons can also suffer Inverse Compton (IC) losses from interactions 
with X-ray and UV photons from the disk. 
To estimate the X-ray photon number density, we take into account 
that the blobs are optically thin \cite{Ghisellini04} so X-ray 
photons escape freely, and their density is
$n_{\rm X}\simeq \alpha_{\rm X} L_{Edd}/4\pi (3r_s)^2 c E_{\rm X}$, or
%
\begin{eqnarray}
n_{\rm X} &\approx & 
\frac{\alpha_{\rm X} L_{Edd} t_{\rm cross}}{(4/3)\pi r_s^3 E_{\rm X}} 
\nonumber \\
&\approx & 5.3 \times 10^{13} \; 
\alpha_{\rm X,0.01} L_{Edd,8} \; E_{\rm X,keV}^{-1}
\;\;\; {\rm cm^{-3}} 
\label{eq:Xdensity}
\end{eqnarray}
where $E_{\rm X}=1\; E_{\rm X,keV}\; {\rm keV}$ is the typical energy 
of the emitted X-rays.

To estimate the density of UV photons from the accretion disk, we 
consider thermal emission in the inner disk between
$\approx 3r_s$ and $7.5r_s$. This portion of the disk emits
an UV luminosity $\alpha_{\rm UV}L_{Edd}$. Observationally
the X-ray luminosity is of the order of 10-50\% the luminosity
in the optical-UV \cite{Walter93}. We then take  
$\alpha_{\rm UV}\approx 5 \alpha_{\rm X} \approx 0.05 \; \alpha_{\rm UV,0.05}$. 
The equilibrium temperature (energy) of the UV photons can
be approximately estimated assuming the emission from the disk
follows a black-body spectrum, yielding:
\begin{eqnarray}
E_{\rm UV} &\approx & k_B \left(\frac{\alpha_{\rm UV}L_{Edd}} 
{\sigma A_{\rm disk}}\right)^{1/4} \nonumber \\
&\approx & 8.3 \; (\alpha_{\rm UV,0.05} L_{Edd,8})^{1/4} \;\;\; {\rm eV}
\end{eqnarray}
where $k_B$ and $\sigma$ are Boltzmann's and 
Stefan-Boltzmann's constant respectively,
and $A_{\rm disk}$ is the area of the inner disk.
The number density of UV photons is then,
\begin{eqnarray}
n_{\rm UV} &\approx & \frac{\alpha_{\rm UV}L_{Edd}}
{A_{\rm disk} c E_{\rm UV}}\nonumber \\
&\approx & 1.2 \times 10^{16} \; (\alpha_{\rm UV,0.5} L_{Edd,8})^{3/4} 
\;\;\; {\rm cm^{-3}}
\label{eq:UVdensity}
\end{eqnarray}
and we get that $n_{\rm UV}\approx 200\; n_{\rm X}$.

\begin{figure}
\centerline{\includegraphics[width=8.cm]{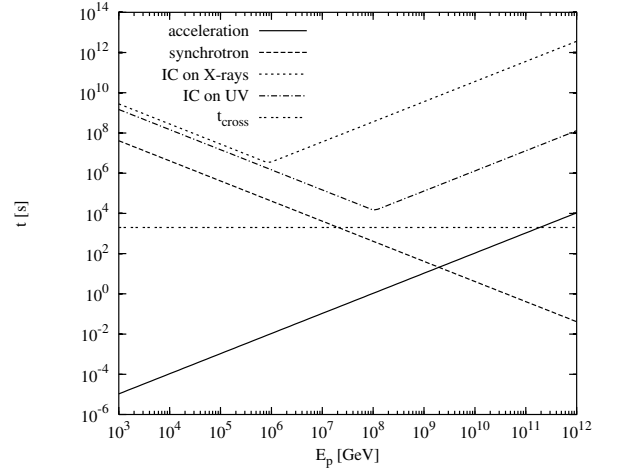}}
\caption{Proton cooling times 
as a function of proton energy for different energy loss processes,
namely, synchrotron and
inverse Compton scattering on both X-rays and UV photons.
Also shown are
the proton acceleration time, and the blob-blob crossing time.}
\label{fig:acceleration}
\end{figure}

The inverse Compton scattering of protons on X-rays and UV photons
is in the Thomson limit for energies below $E_p^{\rm IC}$, where
\begin{eqnarray}
E_p^{\rm IC} & << &  \frac{m_p^2}{E_{\rm X, UV}} \nonumber \\
&\approx & 
\cases{8.8 \times 10^5 \; E_{\rm X, keV}^{-1} 
 & \mbox{GeV \hspace*{0.3cm} (X)} \nonumber \cr  
1.1 \times 10^8 \; (\alpha_{\rm UV,0.05} L_{Edd,8})^{-1/4} 
 & \mbox{GeV \hspace*{0.3cm} (UV)}  } 
\label{eq:Th-limit}
\end{eqnarray}
The IC cooling times on X-ray and UV photons  are then,
in the Thomson (Th) limit
%
\begin{eqnarray}
t_p^{\rm IC,Th} &\approx& \frac{m_p}
{\sigma_{\rm Th}(m_e/m_p)^2 c \gamma_p E_{\rm X,UV} \;n_{\rm X,UV}} \nonumber \\
&\approx &\cases{   
3.0 \times 10^{12} \; 
(\alpha_{\rm X,0.01} L_{Edd,8})^{-1} \gamma_p^{-1}\;\; & 
\mbox{\hspace*{-0.4cm} s \hspace*{0.2cm} (X)} \cr
1.6 \times 10^{12} \; 
(\alpha_{\rm UV,0.05} L_{Edd,8})^{-1} \gamma_p^{-1}\;\; & 
\mbox{\hspace*{-0.4cm} s \hspace*{0.2cm} (UV)}
} \nonumber 
\end{eqnarray}
and in the Klein-Nishina (KN) ($E_p>>E_p^{\rm IC}$) limit
\begin{eqnarray}
t_p^{\rm IC,KN} &\approx& 
\frac{\gamma_p (E_\gamma/m_p)}{\sigma_{\rm Th} (m_e/m_p)^2 c \;n_{\rm X,UV}} 
\nonumber \\
&\approx &\cases{
3.4 \;
(\alpha_{\rm X,0.01} L_{Edd,8})^{-1} E_{\rm X,keV}^{-2} \gamma_p 
& \mbox{\hspace*{-0.2cm} s \hspace*{0.2cm} (X)} \cr
1.2 \times 10^{-4} \; 
(\alpha_{\rm UV,0.05} L_{Edd,8})^{-1/2} \gamma_p 
& \mbox{\hspace*{-0.2cm} s \hspace*{0.2cm} (UV)} 
} \nonumber 
\end{eqnarray}
In Fig. \ref{fig:acceleration} we show the IC cooling times on 
X-rays and UV photons along with the synchrotron cooling and
the proton acceleration times as a function of proton energy. 
Also shown is the dynamic (blob-blob crossing) time. It  
can be seen that IC losses and adiabatic losses (given by the crossing 
time) do not impose any constraints on the maximum proton energy,
which is dominantly restricted by the synchrotron energy losses, and 
is given in Eq. (\ref{eq:Epmax}).  

\section{\label{interactions}Proton interactions}

High energy protons accelerated in the blob-blob collision shock
can interact with X-rays produced in the shocks, UV photons in 
the disk or cold (unaccelerated) protons in the blobs. The 
proton threshold energy for photomeson production on X-rays and 
UV photons at the $\Delta$-resonance 
is given by $E_p E_\gamma\approx 0.3\;{\rm GeV^2}$ which
yields,
\begin{equation}
E_{p,th}^{\rm X} \approx  3.0 \times 10^5 \; E_{\rm X,keV}^{-1} 
\;\;\mbox{GeV}
\label{eq:EpXth}
\end{equation}
for photomeson production on X-rays, and
\begin{equation} 
E_{p,th}^{\rm UV} \approx 3.6 \times 10^7 \; 
(\alpha_{\rm UV,0.05} L_{Edd,8})^{-1/4} 
\;\;\mbox{GeV} 
\label{eq:EpUVth}
\end{equation}
for interactions on UV photons. 

The threshold for inelastic $pp\rightarrow pn\pi^+$ interactions 
in which neutrinos can be produced is much smaller
and is given by,
\begin{equation}
E_{p,th}^p=\frac{(m_p + m_n + m_\pi)^2-2 m_p^2}{2 m_p}
\approx 1.23 \; {\rm GeV}
\end{equation}

The optical depth for photomeson interactions on X-rays, from 
Eq. (\ref{eq:Xdensity}), is 
\begin{equation}
\tau_{p\gamma}^{\rm X}\approx r_s (\sigma_{p\gamma} \rightarrow\Delta) 
n_{\rm X} 
\approx 7.8 \times 10^{-1} \; 
\alpha_{\rm X,0.01} L_{Edd,8} E_{\rm X,keV}^{-1} 
\label{eq:Xdepth}
\end{equation}
and the corresponding optical depth on UV photons, from 
Eq. (\ref{eq:UVdensity}), is,
\begin{equation}
\tau_{p\gamma}^{\rm UV}\approx r_s (\sigma_{p\gamma} \rightarrow\Delta) 
n_{\rm UV} 
\approx 1.8 \times 10^2 \; (\alpha_{\rm UV,0.05} L_{Edd,8})^{3/4} 
\label{eq:UVdepth}
\end{equation}
where $\sigma_{p\gamma\rightarrow\Delta}\approx 5 \times 10^{-28}\;
{\rm cm^2}$ is the photomeson production cross section at the 
$\Delta$-resonance. 

Protons can also interact with cold (unaccelerated) protons in the blob. 
We assume that a fraction $\xi_p<1$ of the protons are Fermi accelerated
in the shocks produced in the blob-blob collisions. The $pp$ collisions 
should dominate below $E_{p\gamma,th}^{\rm X}$, the energy threshold 
for photomeson production on X-rays. In this energy range the mean 
total cross-section for $pp$ interactions is 
$\langle\sigma_{pp}\rangle\approx 6 \times 10^{-26}\; {\rm cm^2}$,
roughly two orders of magnitude larger than the photomeson production 
cross section and the density of proton targets is typically 
$10^3$ times smaller than the X-ray density, 
resulting from Eq. (\ref{eq:pdensity}) in a smaller optical depth
for the $L_X\approx 10^{44}\; {\rm erg\; s^{-1}}$ adopted here,  
\begin{eqnarray}
\langle\tau_{pp}\rangle &\approx & 
\langle\sigma_{pp}\rangle r_s (1-\xi_p) n_p \nonumber \\
&\approx & 1.5 \times 10^{-2} \; (1-\xi_p) 
\alpha_{\rm X,0.01} L_{Edd,8} \; \eta_{0.08}^{-1} 
\end{eqnarray} 
This is the optical thickness for accelerated protons encountering 
blob thermal nucleons. In addition, accelerated  protons emitted in some
directions will also collide with inner disk thermal nucleons. If the
disk inner edge is at $3 r_s$ and has a thickness $3r_s$, a fraction 
$2\pi (3 r_s)^2/ (4\pi (3 r_s)^2) \sim (1/2)$ will collide with nucleons
in the inside wall of the accretion disk, for which $\tau_{pp}$ is larger
due to the larger path length available. The density at inner edge 
is comparable to the blob density, and varies with radius, while the
effective path length depends on the incidence angle. We adopt as an 
estimate for the average proton incident on the disk a value of
$\tau_{pp}\sim 10$ times larger than in a blob.

Inverse Compton scattering of protons on X-rays or UV photons does not
affect the $p\gamma$ scattering rate, since the optical depth for 
IC scattering $\tau_{\rm IC}< 10^{-1}\; \tau_{p\gamma}$ for all 
target photons and in both the Thomson and Klein-Nishina regimes.

The ratio of IC to $pp$ optical depth in the Thomson regime 
($E_p<E_p^{\rm IC}$) is 
\begin{eqnarray}
& & \frac{\tau_{\rm IC}}{\langle\tau_{pp}\rangle} = 
\frac{\sigma_{\rm Th} \; n_{\rm X,UV} \; (m_e/m_p)^2}
{\langle\sigma_{pp}\rangle \; (1-\xi_p) n_p} \\
&\approx &\cases{
2.0 \times 10^{-2} \, 
(1-\xi_p)^{-1} E_{\rm X,keV}^{-1} & \mbox{(X)} \cr 
4.6 \, (1-\xi_p)^{-1} 
\alpha_{\rm X,0.01}^{-1} \alpha_{\rm UV,0.05}^{3/4}
L_{Edd,8}^{-1/4} & \mbox{(UV)} } \nonumber
\label{eq:ICppratio}
\end{eqnarray}
and hence $pp$ interactions which are dominant
below the energy threshold for photomeson 
production on X-rays, are suppressed by IC scattering on UV photons 
for the $L_X\approx 10^{44}\; {\rm erg\; s^{-1}}$ and 
$L_{\rm UV}\approx 5\times 10^{44}\; {\rm erg\; s^{-1}}$ of 
the nominal radio-quiet quasar. In the Klein-Nishina regime 
$\tau_{\rm IC}/\langle \tau_{pp}\rangle$ is lower than 
above, however in this regime $p\gamma$ interactions dominate 
as we will see below. 

\subsection{\label{subsec:power-law}
$p\gamma$ interactions on a power-law X-ray spectrum}

High energy neutrinos are produced in $p\gamma$ interactions
with both X-rays and UV photons above the corresponding energy 
thresholds $E_{p\gamma,th}^{\rm X}\approx 3.0 \times 10^5$ GeV 
and $E_{p\gamma,th}^{\rm UV}\approx 3.6 \times 10^7$ GeV respectively.
However, the X-ray spectra of AGN, including the radio-quiet, are 
typically of power-law form, and hence protons at energies below 
$E_{p\gamma,th}^{\rm X}$ can also produce delta resonances by 
interacting with photons from the high energy tail of the X-ray 
spectrum. The same is true for $p\gamma$ interactions with UV photons 
in which low energy protons can photoproduce on photons from the
high energy tail of the blackbody photon distribution. 
As a first approximation, we will neglect the tail of the blackbody
distribution, and assume that the UV spectrum is monoenergetic. 
However, we will take into account that the X-ray spectrum follows a 
power-law given by $dN_\gamma/dE_\gamma \propto E_\gamma^{-\delta}$.
As a consequence the optical depth to $p$-X-rays depends on the 
spectral index $\delta$ of the photon spectrum, as well as on 
proton energy. This is due to the fact that as $E_p$ increases, 
protons interact with an increasing number of lower energy photons. 
For a steep X-ray spectrum, the bulk of the X-ray photons are
concentrated in the low energy region of the power-law distribution, 
and this increases the optical depth to $p$-X-ray interactions for 
high energy protons.      
For a photon number density spectrum $\propto E^{-\delta}$,
the energy dependent optical depth is given by,
\begin{equation}
\tau_{p\gamma}^{\rm X}(E_\gamma) \simeq 
r_s (\sigma_{pX \rightarrow \Delta}) E_\gamma \frac{dn_\gamma}{dE_\gamma}
\propto E_\gamma^{1-\delta}  
\label{eq:tauegamma}
\end{equation}
Using the threshold relation $E_p E_\gamma \approx 0.3 \; {\rm GeV^2}$
Eq. (\ref{eq:tauegamma}) turns into,
\begin{eqnarray}
\tau_{p\gamma}^{\rm X}(E_p) &\simeq & 
\tau_{p\gamma}^{\rm X}(E_\gamma=1 \; {\rm keV}) 
\left(\frac{E_p}{E_{p,th}^X}\right)^{\delta-1} \nonumber \\ 
&\approx &  7.8 \times 10^{-1} \; (\alpha_{\rm X,0.01} L_{Edd,8}) 
\left(\frac{E_p}{E_{p,th}^X}\right)^{\delta-1}
\label{eq:taupgamma_Ep}
\end{eqnarray}
where $\tau_{p\gamma}^{\rm X}(E_\gamma=1 \; {\rm keV})
:=\tau_{p\gamma}^{\rm X,{\rm keV}}$ is the optical
depth given in Eq. (\ref{eq:Xdepth}). The factor 
$({E_p}/{E_{p,th}^X})^{(\delta-1)}$ accounts for the increasing 
optical depth as $E_p$ increases (note that this is only for 
$\delta>1$, which is observationally the case).

\begin{figure}
\centerline{\includegraphics[width=8.5cm]{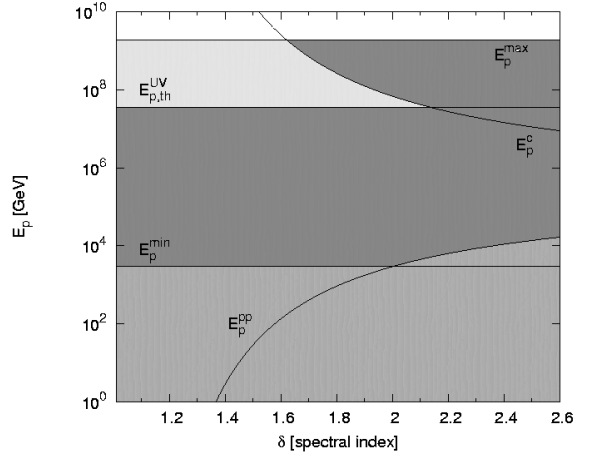}}
\vspace*{-0.7cm}
\caption{
Several energy scales relevant for proton interactions for
the set of parameters of the nominal radio-quiet quasar:
$M_{\rm BH}=10^8 M_\odot$, $\alpha_{X,0.01}=1$, $\alpha_{UV,0.05}=1$.
$E_p^{max}$ is the maximum energy for proton acceleration. 
$E_{p,th}^{\rm UV}$ is the proton energy threshold for photomeson
production on UV photons. $E_p^{min}$ is the lowest energy of a proton 
that can interact through the $\Delta-$resonance 
with an X-ray assuming the maximum X-ray energy is $E_\gamma^{max}=100$ keV.  
The upper curve is $E_p^c$, 
the energy above which the optical depth
for $p\gamma$ interactions on X-rays is larger than the corresponding
optical depth on UV photons. The lower curve is $E_p^{pp}$ the energy below
which the $pp$ optical depth is larger than the $p\gamma$ optical
depth on X-rays. $p\gamma$ interactions on X-rays are dominant in 
the darkest grey shaded parameter space region $(E_p,\delta)$. 
$p\gamma$ interactions on UV photons dominate in the lightest gray
shaded area, whereas $pp$ interactions are more important in the
medium gray shaded region at the bottom of the plot.}
\label{fig:pinteractions}
\end{figure}

From the discussion above, it follows that (for $\delta>1$) 
there is a proton energy $E_p^c$  above which photoproduction on 
X-rays dominates over photoproduction on UV photons. The energy
$E_p^c$ is given by the condition $\tau_{p\gamma}^{\rm X}(E_p^c)=
\tau_{p\gamma}^{\rm UV}$. Solving for $E_p^c$ we have
\begin{eqnarray}
& & E_p^c = \left[\frac{\tau_{p\gamma}^{\rm UV}}
{\tau_{p\gamma}^{\rm X,{\rm keV}}}\right]^{1/(\delta-1)} 
E_{p,th}^X \nonumber \\
&\approx & [229.6\;  \alpha_{\rm X,0.01}^{-1} \; \alpha_{\rm UV,0.05}^{3/4} \; 
L_{Edd,8}^{-1/4}]^{1/(\delta-1)} \; E_{p,th}^X
\label{eq:Epcut}
\end{eqnarray}
The predominant target for $p\gamma$ interactions is determined by 
the numerical value of $E_p^c$ and whether it is larger or 
smaller than $E_{p,th}^{\rm UV}$, the energy above which $p$'s 
photoproduce on UV photons, as well as by its relative value with 
respect to  $E_p^{max}$, the maximum energy of the accelerated 
protons. $E_p^c$ depends on the X-ray spectral index $\delta$. 
It follows that:
\begin{itemize}

\item{If $\delta \geq \delta_c\approx 2.13$ then 
$E_p^c \leq E_{p,th}^{\rm UV}$, implying that proton interactions 
with X-rays photons always dominate over interactions with UV photons.}

\item{If $1 < \delta \leq \delta_l\approx 1.61$ then $E_p^c \geq 
E_p^{max}$ and interactions with X-rays dominate up to 
$E_{p\gamma,th}^{\rm UV}$, while photoproduction on UV photons 
dominates above $E_{p\gamma,th}^{\rm UV}$ }

\item{If $\delta_l < \delta < \delta_c$ there is a small proton energy
range above $E_{p\gamma,th}^{\rm UV}$ but below $E_p^{max}$ in which 
photoproduction on UV photons dominates, whereas interactions with X-rays
are more important outside it.}

\end{itemize}
The numerical value of $\delta_c$ is obtained from the condition 
$E_p^c(\delta_c)=E_{p,th}^{\rm UV}$ whereas $\delta_l$ 
is obtained from the equation $E_p^c(\delta_l)=E_p^{max}$.
They depend on the particular values of  
$L_{\rm X}$ and $L_{\rm UV}$.
The situation is illustrated in Fig. \ref{fig:pinteractions}, 
where $E_p^c$ is plotted as a function of $\delta$, along with 
$E_p^{max}$ and $E_{p\gamma,th}^{\rm UV}$. The dark grey shaded 
area represents the region in the space parameter of 
$(E_p,\delta)$ where interactions on X-rays are dominant, while
photoproduction on UV photons is more important in the light
gray shaded area. Interactions above $E_p^{max}$ (unshaded region) 
are not possible, since protons do not get accelerated to 
$E_p>E_p^{max}$.  

For the $pp$ interactions, similarly to the discussion above, 
there is a proton energy $E_p^{pp}$ below which $pp$ interactions 
are more important than interactions with X-rays, given by the 
condition $\tau_{p\gamma}^{\rm X}(E_p^{pp})=\langle\tau_{pp}\rangle$. 
Solving for $E_p^{pp}$ we get:
\begin{eqnarray}
& & E_p^{pp} = \left[\frac{\langle\tau_{pp}\rangle}
{\tau_{p\gamma}^{\rm X,{\rm keV}}}\right]^{1/(\delta-1)} \; 
E_{p,th}^X \nonumber \\
&\approx &[9.8 \times 10^{-3}]^{1/(\delta-1)}\; E_{p,th}^X 
\end{eqnarray}
There is also another important energy scale, 
$E_{\gamma}^{max}$ which is the photon energy above which the 
source runs out of energetic enough X-rays to interact with 
protons of energy below $E_p^{min}$ given by the photoproduction 
threshold condition: 
$E_p^{min} E_{\gamma}^{max}\approx 0.3 \; {\rm GeV^2}$.  
Observationally, most radio-quiet AGN spectra show 
an X-ray cut-off at about 100-150 keV. We will 
adopt $E_{\gamma}^{max}=100$ keV in this work.

As in the case of $p\gamma$ interactions, $E_p^{pp}$ depends on the
index of the X-ray spectrum. For a steep enough spectrum there are
not enough high energy photons to interact with low energy protons 
and then $pp$ interactions dominate below a proton energy larger 
than $E_p^{min}$. This occurs for $\delta\geq \delta_p \sim 2.00$ given by 
$E_p^{pp}(\delta_p)=E_p^{min}$, using $\xi_p=0.5$ and the 
standard set of luminosities of the nominal radio-quiet AGN. 
This is also shown in Fig. \ref{fig:pinteractions} where 
$E_p^{pp}$ is plotted along with $E_p^{min}$ assuming 
$E_{\gamma}^{max}= 100$ keV. Thus, $pp$ interactions  
dominate in the medium gray region at the bottom of 
Fig. \ref{fig:pinteractions} if the $p$ flux is not suppressed
by IC scattering, which depends on the luminosity of the source. 

\section{\label{neutrinos}Neutrino production and detection}

High energy neutrinos are produced through secondary charged $\pi$ decay
photoproduced in $p\gamma\rightarrow\Delta$ and 
$n\gamma\rightarrow\Delta$ interactions. Each 
of the 4 leptons produced in the charged pion decay chain 
$\pi^\pm \rightarrow \mu \nu_\mu \rightarrow e\nu_e
\nu_\mu \bar\nu_\mu$ carry about 1/4 of the initial pion energy.

Neutrinos can also be produced through pion and muon decay produced 
in $pp$ interactions in the shocks. In fact $pp$ interactions are 
dominant at low proton energies (medium-grey shaded region at the 
bottom of Fig.\ref{fig:pinteractions}). The typical neutrino energy
is 1/4 of the initial pion energy, and this is 
determined by the pion multiplicity in the $pp$ collisions.  

Due to the high magnetic field and dense photon environment where the 
$\pi^+$ and $\mu^+$ are produced, they might lose their energies 
through synchrotron and inverse Compton scattering before decaying
to neutrinos. 

The synchrotron cooling time for pions and muons is calculated 
using Eq. (\ref{eq:synchrotron}). The charged pion and muon decay times
are $\tau_\pi^{dec}\approx 2.6 \times 10^{-8}\;\gamma_\pi$ s, and
$\tau_\mu^{dec}\approx 2.2 \times 10^{-6}\;\gamma_\mu$ s. Equating
the synchrotron cooling time and the decay time, we get the 
synchrotron break energies above which neutrino production is 
suppressed by a factor $\propto E_\nu^{-2}$ due to energy losses:
\begin{equation}
E_\pi^{sb} \approx 1.0 \times 10^7 \; 
(\alpha_{\rm X,0.01} L_{Edd,8})^{-1/2} 
\epsilon_{B,0.3}^{-1/2} \;\; {\rm GeV}
\label{eq:Ebreakpi}
\end{equation}
for pions, and,
\begin{equation}
E_\mu^{sb} \approx 5.6 \times 10^5 \;
(\alpha_{\rm X,0.01} L_{Edd,8})^{-1/2} 
\epsilon_{B,0.3}^{-1/2} \;\; {\rm GeV}
\label{eq:Ebreakmu}
\end{equation}
for muons.  

Inverse Compton scattering on both X-rays and UV photons does not contribute
significantly to the pion and muon energy loss. For all photon targets and
in both the Thomson and Klein-Nishina regimes, the inverse Compton cooling
time is at least $\approx 10$ times larger than the decay times (and
much larger than this $\approx 10^2 - 10^7$ in most cases) and 
hence we ignore pion and muon IC scattering 
energy losses in the calculation of the neutrino flux.  

\subsection{Neutrino flux from the nominal radio-quiet quasar}

Protons are accelerated by the Fermi mechanism 
in the blob-blob shocks up to energies 
$E_p^{max}\approx 10^9$ GeV, following an energy spectrum 
$\propto E_p^{-2}$. The energy per unit time given 
to protons producing neutrinos is,
\begin{equation}
\int_{E_{p,th}^p}^{E_p^{max}} \; E_p \frac{d^2 N_p}{dE_p dt}\; dE_p =
\xi_p \; \alpha_{\rm X} L_{Edd}
\end{equation}
Taking $d^2 N_p/dE_p dt = K_p/E_p^2$ we get for $K_p$,
\begin{equation}
K_p\approx 3.2 \times 10^{45} \; 
\frac{\xi_p \; \alpha_{\rm X,0.01} L_{Edd,8}}{\ln (E_p^{max}/E_{p,th}^p)}\;\;
{\rm GeV\;s^{-1}}
\end{equation}
The proton fluence per unit time is,
\begin{eqnarray}
F_p &=& \frac{1}{4\pi D_L^2} \; 
E_p^2 \frac{d^2 N}{dE_p dt} \\ 
&\approx & 1.6 \times 10^{-5} \; \frac{\xi_p \; \alpha_{\rm X,0.01} L_{Edd,8}}
{D_{L,1}^2 \; \ln (E_p^{max}/E_{p,th}^p)} \;\; 
{\rm \frac{GeV}{cm^2 \; s}} \nonumber
\label{eq:pfluence}
\end{eqnarray}
for a source at a luminosity distance $D_L=1\; D_{L,1}$ Mpc. 

\subsubsection{Neutrino flux from $p\gamma$ interactions}

The neutrino flux for each neutrino flavor is given by, 
\begin{eqnarray}
\Phi_\nu^{p\gamma} &=& \frac{1}{4} f_\pi F_p \\
&\times & 
\cases{
E_\nu^{-2} \;
& \mbox{;    $E_{\nu,th}^p < E_\nu < E_\nu^{sb}$} \cr 
E_\nu^{-2} \; 
(E_\nu^{sb}/E_\nu)^2 \; 
& \mbox{;    $E_\nu^{sb} < E_\nu < E_\nu^{max}$}
} \nonumber
\label{eq:nufluxquasar}
\end{eqnarray}
where $E_\nu^{sb}\approx (1/4) \; E_\pi^{sb}$ for
muon neutrinos produced in $\pi$ decays, and
$E_\nu^{sb}\approx (1/4) \; E_\mu^{sb}$ for electron
and muon neutrinos produced in the decay of $\mu$'s.
$E_\pi^{sb}$ and $E_\mu^{sb}$ are given in 
Eqs. (\ref{eq:Ebreakpi}) and (\ref{eq:Ebreakmu}) respectively,
$E_{\nu,th}^p=0.05 \; E_{p,th}^p$ and 
$E_\nu^{max}=0.05 \; E_p^{max}$.

$f_\pi$ is the fraction of proton energy given to pions.
As discussed in subsection \ref{subsec:power-law}, $f_\pi$
depends on proton energy (and hence on neutrino energy)
and on $\delta$ the spectral index of the X-ray energy distribution,
which determine the dominant photon target (X-rays or UV photons). 
For interactions on X-rays $f_\pi$ is given by,  
\begin{equation}
f_\pi^{\rm X}={\rm min}\left[1,\;\langle x_{p\rightarrow \pi} \rangle
\tau_{p\gamma}^{\rm X,keV}
\left(\frac{E_\nu}{E_{\nu,th}^{\rm X}}\right)^{\delta-1}\right]
\end{equation}
where $\langle x_{p\rightarrow \pi}\rangle\approx 0.2$ is the
average inelasticity in a $p\gamma$ interaction, 
$\tau_{p\gamma}^{\rm X,keV}$ is given in Eq. (\ref{eq:Xdepth}) and
${E_{\nu,th}^{\rm X}}=0.05\;{E_{p\gamma,th}^{\rm X}}$ given in 
Eq. (\ref{eq:EpXth}). 
For interactions on UV photons,
\begin{equation}
f_\pi^{\rm UV}={\rm min}(1,\;\langle x_{p\rightarrow \pi} \rangle
\tau_{p\gamma}^{\rm UV})
\end{equation}
where $\tau_{p\gamma}^{\rm UV}$ is given in Eq. (\ref{eq:UVdepth}).

The dependence of $f_\pi$ on $\delta$ and $E_\nu$ 
can be summarized as follows,

\begin{itemize}

\item{If $\delta \geq \delta_c$, then: 
\begin{equation}
f_\pi=f_\pi^{\rm X}; \;\;\; 
E_{\nu,th}^p < E_\nu < E_\nu^{max}
\end{equation}
}
\item{If $1 < \delta \leq \delta_l$, then:
\begin{equation}
f_\pi=\cases{
f_\pi^{\rm X}; \;\;\; E_\nu < E_{\nu,th}^{\rm UV} \nonumber \cr
f_\pi^{\rm UV}; \; E_{\nu,th}^{\rm UV} < E_\nu < E_{\nu}^c \cr 
f_\pi^{\rm X}; \;\;\; E_\nu > E_{\nu}^c \nonumber 
}
\end{equation}
where $E_{\nu,th}^{\rm UV}\approx 0.05 \; E_{p\gamma,th}^{\rm UV}$
given in Eq. (\ref{eq:EpUVth}), and 
$E_{\nu}^c \approx 0.05 \; E_p^c$ given in Eq. (\ref{eq:Epcut}).
}
\item{If $\delta_l < \delta < \delta_c$, then:
\begin{equation}
f_\pi=\cases{
f_\pi^{\rm X}; \;\;\; E_\nu < E_{\nu,th}^{\rm UV} \cr
f_\pi^{\rm UV}; \;\;\; E_\nu > E_{\nu,th}^{\rm UV} 
}
\end{equation}
}

\end{itemize}

\subsubsection{Neutrino flux from $pp$ interactions}

The neutrino flux from $pp$ interactions is \cite{Razzaque03} 

\begin{eqnarray}
\Phi_\nu^{pp} &=& 
\frac{1}{2}\int\;f_{pp,1}\;M_\nu(E_p)\;
\frac{F_p}{E_p^2}\;dE_p \\
&+& \frac{1}{2}\int\;f_{pp,2}\;M_\nu(E_p)\;
\frac{F_p}{E_p^2}\;dE_p
\end{eqnarray}
for $E_p < E_{p,th}^p$, where 
$f_{pp,1}={\rm min}(1,\langle \tau_{pp} \rangle)$,
and $f_{pp,2}={\rm min}(1,10\langle \tau_{pp} \rangle)$ 
accounting for the larger optical depth  
in the inner wall of the accretion disk
where $\sim 1/2$ of the accelerated protons interact 
(see Section \ref{interactions}). $M_\nu(E_p)$ is
the neutrino multiplicity through pion decay in the $pp$ interactions
which can be parameterized as \cite{Razzaque02},
\begin{eqnarray}
M_{\nu}(E_p) &=& \frac{7}{4} \left(
\frac{E_{\nu}}{\rm GeV} \right)^{-1} \left[ \frac{1}{2}{\rm ln}\left(
\frac{10^{11}\, {\rm GeV}}{E_p} \right) \right]^{-1} \nonumber \\
&\times & \Theta \left( \frac{1}{4} \frac{m_{\pi}}{\rm GeV}
\gamma_{\rm CM} \leq \frac{E_{\nu}}{\rm GeV} \leq \frac{1}{4
}\frac{E_p}{\rm GeV} \right)
\label{pp-numult}
\end{eqnarray}
for each neutrino flavor.

\begin{figure}
\centerline{\includegraphics[width=8.5cm]{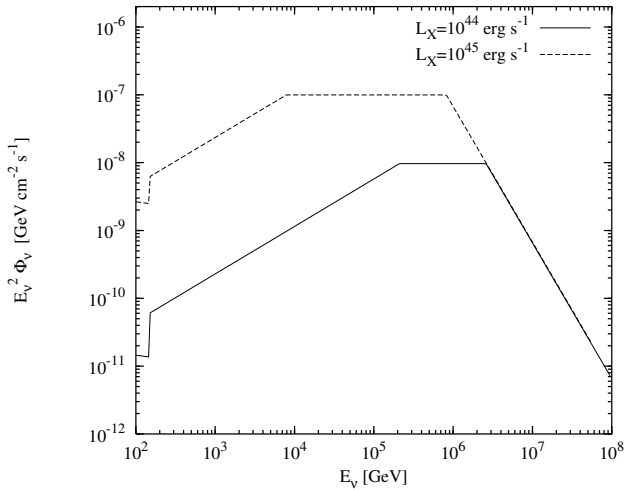}}
\caption{Solid line: Neutrino flux reaching Earth for the nominal radio-quiet
quasar: $M_{\rm BH}=10^8 M_\odot$,
$\alpha_{\rm X}=0.01$ ($L_{X}\approx 10^{44}\;{\rm erg\;s^{-1}}$) 
and $\alpha_{\rm UV}=0.05$ at a luminosity
distance $D_L=20$ Mpc, $(1+z)\approx 1$. Dashed line: Neutrino flux for a set 
of parameters $M_{\rm BH}=10^8 M_\odot$, 
$\alpha_{\rm X}=0.1$ ($L_{X}\approx 10^{45}\;{\rm erg\;s^{-1}}$)
and $\alpha_{\rm UV}=0.5$ at $D_L=20$ Mpc. 
Only $\nu_\mu$ from pion decay produced
in $pp$ and $p\gamma$ interactions are shown in 
the figure.}
\label{fig:nuflux_single}
\end{figure}

Figure \ref{fig:nuflux_single} shows the neutrino flux 
produced in $pp$ and $p\gamma$ interactions,
for the nominal radio-quiet quasar at a distance $D_L=20$ Mpc. The quasar
X-ray spectrum is fairly universal, typically being $E_\gamma^{-\delta}$
with $\delta=1.7$ \cite{Turner89}. For this spectral index and for the 
nominal radio-quiet quasar, interactions with unaccelerated protons, 
X-rays and UV photons contribute to the neutrino flux 
(see Fig. \ref{fig:pinteractions}). Below the proton threshold for 
interaction on X-rays, $pp$ interactions dominate neutrino production, 
however $pp$ collisions are largely suppressed due to the more dominant
inverse Compton scattering on UV photons (see Eq. \ref{eq:ICppratio}). 
This behavior continues until the protons have enough energy to 
interact with the high energy tail of the X-ray photon spectrum. 
This effect accounts for the rapid increase in the neutrino flux around
$\approx 200$ GeV seen in Fig. \ref{fig:nuflux_single}. 
After that the flux
is slightly flatter than the expected $E_\nu^{-2}$ power law due to the 
$E_\nu^{\delta-1}=E_\nu^{0.7}$ factor in the $\tau_{p\gamma}^{\rm X}$
optical depth. As energy increases $f_\pi$ reaches 1 and the neutrino
flux goes as $E_\nu^{-2}$, 
before synchrotron losses by 
pions become important  
making the neutrino flux to fall as $E_\nu^{-4}$. 

The normalization as well as  
the shape of the neutrino flux changes with luminosity. This is
illustrated in Fig. \ref{fig:nuflux_single} for a radio-quiet 
quasar of $L_X\approx 10^{45}\;{\rm erg\;s^{-1}}$ and 
$M_{\rm BH}=10^8\;M_\odot$. For larger $L_{\rm X}$ 
the optical depth to $pX-$rays interactions reaches 1 at 
a smaller proton (and hence neutrino) energy due to the denser
photon target, and as a consequence 
the $E_\nu^{-2}$ behavior of the spectrum 
starts at a lower energy than in the nominal AGN. 
Besides, since for fixed black hole mass, 
the pion synchrotron break energies scales
as $L_{\rm X}^{-1/2}$ (see Eq. (\ref{eq:Ebreakpi})) the 
flux starts to fall as $E_\nu^{-4}$ at a lower energy than for the 
lower luminosity AGN. 

\subsection{Diffuse neutrino flux}

The cumulative neutrino flux from all radio-quiet AGNs 
in the Universe is obtained by convolution of the observed 
point source neutrino 
flux with the luminosity function taking into 
account its cosmological evolution, 
\begin{eqnarray}
& & \Phi_{\nu,{\rm ob}}^{\rm diff} (E_{\nu,{\rm ob}}) = \nonumber \\ 
& & \frac{1}{4\pi}
\int_{L_{\rm min}}^{L_{\rm max}} dL_{\rm X} \; \int_0^{z_{\rm max}}
dz \; \frac{dn_0}{dL_{\rm X}} \; f(z) \; \frac{dV}{dz}\; 
\Phi_{\nu,{\rm ob}} \; 
\end{eqnarray}
where $dn_0/dL_{\rm X}$ describes the present day X-ray luminosity function 
of the sources and $f(z)$ is its cosmological evolution. We have used
the broken power-law luminosity and evolution functions of reference 
\cite{Boyle93} (model I). In a Friedmann-Robertson-Walker universe
the comoving volume element is 
$dV/dz = 4\pi D_L^2 c\vert dt/dz \vert/(1+z)$ and the derivative
of the cosmic time $t$ with respect to redshift $z$ is
$(dt/dz)^{-1}=-H_0(1+z)\sqrt{(1+\Omega_m z)(1+z)^2-\Omega_\Lambda(2z+z^2)}$ 
where we have used a standard $\Lambda$CDM cosmology with
$\Omega_m=0.3$ and $\Omega_\Lambda=0.7$. 
The neutrino energy in the observer's frame and the source frame 
are related by $E_\nu=E_{\nu,{\rm ob}}(1+z)$, and the luminosity
at the source is $(1+z)^2$ the luminosity observed today. 
In the integration  
we take into account that the shape of the observed individual
neutrino spectrum depends on the luminosity of the 
source as illustrated in Fig. \ref{fig:nuflux_single}. 

We have performed the calculation of the diffuse flux in two 
different ways. First we assume that in all the individual sources
the same fraction of the 
Eddington luminosity ($\alpha_{\rm X}=0.01$) is converted to X-rays,
adjusting the mass of the black hole in order to get the necessary X-ray 
luminosity (variable $M_{\rm BH}$ case). 
As an alternative calculation,
we assume the black hole has the same mass in all sources 
($M_{\rm BH}=10^8\;M_\odot$) and we vary the efficiency of conversion
of Eddington luminosity into X-rays, i.e. we vary $\alpha_{\rm X}$,
adjusting its value so that we get the required $L_{\rm X}$ 
(fixed $M_{\rm BH}$ case).

\begin{figure}
\centerline{\includegraphics[width=8.5cm]{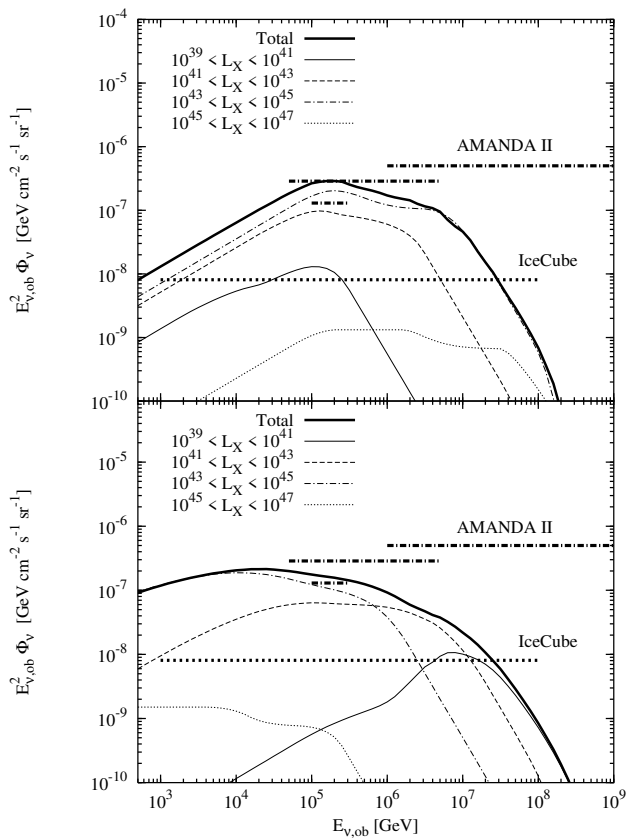}}
\vspace*{-0.7cm}
\caption{ Radio-quiet AGN diffuse $\nu_\mu+\bar\nu_\mu$ flux reaching
Earth (thick solid line). The fact that half of the muon neutrinos 
disappear due to oscillations is accounted for. The contribution to 
the total neutrino flux (thick lines) from radio-quiet AGNs in 
different X-ray luminosity bins (in ${\rm erg\;s^{-1}}$) is shown in 
thin lines. The top panel shows the results of a calculation where
the black hole mass is varied, while the bottom panel shows the 
case where the accretion rate is varied (see text for details). 
Also shown are the latest AMANDA II limits on an $E_\nu^{-2}$ 
diffuse $\nu_\mu + \bar\nu_\mu$ flux which apply in different 
energy ranges (thick dashed-dotted lines) \cite{AMANDAII_limit,AMANDAII_limit2},
and the expected IceCube sensitivity after 1 year of operation 
(for an $E_\nu^{-2}$ diffuse $\nu_\mu+\bar\nu_\mu$ flux at 90$\%$ C.L.)
\cite{icecube} (thick dashed line).}
\label{fig:nuflux_diffuse}
\end{figure}

In Fig. \ref{fig:nuflux_diffuse} we plot the diffuse 
$\nu_\mu + \bar\nu_\mu$ flux as obtained in both calculations. 
The contribution to the total neutrino flux from radio-quiet 
AGNs in different luminosity bins is shown. The less luminous
AGNs, although more abundant, do not contribute much to the total because
the neutrino flux scales with luminosity. The most luminous sources,
although powerful neutrino emitters, are less abundant and contribute
very little to the total flux. 

In the case of fixed black hole mass  
(bottom panel) the dimensions of the accretion disk and black
hole region are also fixed, and as luminosity increases the density 
of protons and X-rays increases linearly with $L_{\rm X}$. 
Proton interactions with X-rays start to become important at 
an increasingly smaller energy as $L_{\rm X}$ increases due to
the denser X-ray target. At the same time, the synchrotron 
break occurs at an increasingly lower energy because the
magnetic field increases as $L_{\rm X}^{1/2}$ (see Eq. (\ref{eq:Bfield})). 
The two effects are visible in Fig. \ref{fig:nuflux_diffuse}. 

In the case of variable $M_{\rm BH}$ (top panel), an increase 
in luminosity for a fixed accretion efficiency is accompanied 
by an increase of the black hole mass, i.e. the accretion disk
and black hole region becomes larger, and the proton
and photon density actually decrease as $M_{\rm BH}^{-1}$. This 
has the effect of decreasing the magnetic field intensity in the 
source and hence the synchrotron break occurs at larger energy
as $L_{\rm X}$ increases. However the optical depth to $p\gamma$
interactions stays the same as $M_{\rm BH}$ increases and as 
a consequence the proton energy at which it reaches 1, i.e. the
energy at which the neutrino flux starts falling as $E_\nu^{-2}$, 
stays the same. The two effects are again visible in the top 
panel of Fig. \ref{fig:nuflux_diffuse}. 

Using Eq. (\ref{eq:taupgamma_Ep}) one sees that for low
luminosity AGNs, $L_{\rm X}<10^{41}\; {\rm erg \;s^{-1}}$,
$\tau_{p\gamma}^{\rm X}<1$ for all proton energies
up to $E_p^{max}$, and protons can escape from the sources
contributing to the observed cosmic ray spectrum.
As a consequence, neutrinos produced in the blob-blob shocks 
in low luminosity AGNs would in principle be affected by the 
Waxman-Bahcall bound:
$E_\nu^2 dN_\nu/dE_\nu < (1-4) \times 10^{-8}\;
{\rm GeV \; cm^{-2} \; s^{-1}}$ \cite{Waxman-Bahcall}.
However, the contribution of sources with
$L_{\rm X}<10^{41}\; {\rm erg \;s^{-1}}$ to the total
neutrino flux is in any case below the Waxman-Bahcall bound,
as can be seen in Fig. \ref{fig:nuflux_diffuse}, while for 
larger luminosities, the neutrinos are produced in regions where 
the optical depth to $p\gamma$ collision is typically $\geq 1$.
Thus the Waxman-Bahcall bound does not apply to the sources 
considered here. 

Also shown in Fig. \ref{fig:nuflux_diffuse} are the latest 
AMANDA II limits on $\nu_\mu + \bar\nu_\mu$ for an 
$E_\nu^{-2}$ diffuse flux \cite{AMANDAII_limit,AMANDAII_limit2} 
which apply in different energy ranges, 
as well as the  expected IceCube sensitivity (90$\%$ C.L.) after 1 year 
of operation for an $E_\nu^{-2}$ flux \cite{icecube}. The energy
range in which the IceCube sensitivity is shown, is for illustrative
purposes only.

It is important to remark that the nominal muon event rates 
(see below), as well as the individual source and diffuse neutrino 
fluxes of figures \ref{fig:nuflux_single} and \ref{fig:nuflux_diffuse} 
are proportional to $\alpha_{bb}$ (the fraction of the X-ray flux in 
radio-quiet AGN which is due to a core component) and to $\xi_p$ 
(the fraction of protons which is accelerated in core shocks of
any origin, attributed here more specifically to blob-blob collisions). 
The values adopted here as an example, $\alpha_{bb}\sim 1$ and 
$\xi_p\sim 0.5$, are reasonable but not well constrained so far.
Additional uncertainties for the diffuse neutrino flux are the
approximate luminosity function, and the fraction of radio-quiet 
AGNs (which may be thought of as an average $\alpha_{bb}$)
for which a core X-ray component requires invoking shocks,
blob collisions, aborted jets or other mechanisms besides an 
accretion disk. The AMANDA II neutrino detector is currently  
sensitive to the diffuse neutrino fluxes predicted here with
the above parameters (see Fig. \ref{fig:nuflux_diffuse}), and 
hence gives important information about the relevant astrophysical
parameters of the aborted jets scenario, namely, that the product
$\alpha_{bb} \times \xi_p$ may be a factor $\lesssim 1/2$ times the nominal 
values adopted here. The larger ${\rm km^3}$ IceCube will be able 
to detect the diffuse fluxes in less than 1 year of operation, or 
otherwise rule out the generic model explored in this paper. 

\subsection{Event rate in a Gigaton telescope}

We estimate here the neutrino event rate from a single nearby 
radio-quiet AGN, taking as an example a distance $D_L\approx 20$ Mpc, 
comparable to that of the Virgo cluster of galaxies.  We assume the 
nominal radio-quiet AGN with parameters: $M_{\rm BH}=10^8 M_\odot$, 
$\alpha_{\rm UV,0.05}=1$, and $\alpha_{\rm X,0.01}=1$, i.e. $L_X\sim
10^{44}$ erg s$^{-1}$. Considering neutrinos from $p\gamma$ interactions,
the energy in neutrinos reaching the Earth per unit area and time from 
a source at this distance, from Eqs. (\ref{eq:pfluence}) and 
(\ref{eq:nufluxquasar}), is 
$F_\nu\approx 10^{-8}\;{\rm GeV \; cm^{-2} \; s^{-1}}$, where we 
have used $\xi_p=0.5$, $f_\pi=1$, $E_p^{max}=2 \times 10^9$ GeV 
and $E_{p,th}^p=1.23$ GeV. The number of neutrinos reaching the Earth
assuming a typical neutrino energy $E_\nu=100\;{\rm TeV}=10^5$ GeV is
$\approx F_\nu/E_\nu\approx 10^{-13}\;{\rm cm^{-2} \; s^{-1}}$. 
At this energy the probability that a neutrino converts into 
a muon within the range of the detector is $\approx 10^{-4}$. 
The number of detectable muons is $\approx 3$ per ${\rm km^2}$ 
per year. This number scales roughly with the luminosity of the 
source and the inverse of the distance squared. The estimate only 
takes into account muon events from $\nu_\mu$'s produced directly 
in $\pi$ decay. We have also made a more detailed estimate that 
convolutes the neutrino flux with the probability of neutrino 
detection through both muon and shower observation, taking into 
account the absorption of the neutrino flux while passing through 
the Earth (see for instance Appendix C in \cite{guetta04}), and 
assuming that after propagation from the source neutrinos have 
oscillated so that the flavor ratio is
$\nu_e:\nu_\mu:\nu_\tau=1:1:1$. We obtain a muon event rate  
$N_\mu\approx 9\;{\rm km^{-2}\;yr^{-1}}$ and a shower rate
$N_{\rm sh}\approx 2\;{\rm km^{-2}\;yr^{-1}}$ for an AGN located 
at a zenith angle $90$ degrees (with respect to the vertical
direction) in a detector such as Icecube at an average depth of 
$\sim$1.8 km water equivalent.
The muon and shower energy threshold we used is 500 GeV. 

For the diffuse fluxes plotted in Fig. \ref{fig:nuflux_diffuse} 
we predict a muon rate 
integrated over the $2\pi$ northern sky (where the background due 
to atmospheric muons is negligible) of the order of a few hundred
events and a shower rate about 3 times smaller, with roughly 
the same rate for the fixed $M_{\rm BH}$ case.  
IceCube can also identify events 
coming from the southern hemisphere despite the large muon background
by selecting events with high enough energy \cite{icecube,alvarez-muniz01}. 
The number of muon events
expected in this model from the $2\pi$ southern sky for an 
energy threshold of $10^6$ GeV is $N_\mu\approx 17 \; {\rm km^{-2} \; yr^{-1}}$ 
for the variable $M_{\rm BH}$ case and 
$N_\mu\approx 8 \; {\rm km^{-2} \; yr^{-1}}$ for the fixed $M_{\rm BH}$ 
case.

\section{\label{conclusions}Summary and conclusions}

We have investigated the high energy neutrino emission from 
radio-quiet AGNs, which is the most numerous class of AGNs.
The neutrino emission in these AGNs, which lack prominent jets, is 
attributed to $p\gamma$ and $pp$ collisions of relativistic protons 
accelerated in shocks occurring near the central black hole, which 
are also responsible for a core component of the observed X-ray 
luminosity. The calculations were performed for a specific model of 
the AGN core in which an abortive jet leads to collisions of blobs 
inside the inner edge of the accretion disk. The results, however, 
should be generic to any model involving particle acceleration in the 
inner region of the accretion flow, e.g. \cite{kazanas86,stecker96}. 
This model allows us to predict spectral details whose features 
depend on physical parameters related to the black hole mass, 
accretion rate and X-ray/UV luminosity, as well as shock physics 
parameters such as proton injection fraction, shock magnetic field 
strengths, etc. 
In particular, the model predicts muon and shower event rates from 
individual AGN, as a function of their observed X-ray/UV luminosity.
The individual source event rates, as well as the corresponding
diffuse fluxes are, for nominal parameters, already 
constrainable with AMANDA II, and should be strongly 
constrainable with the IceCube detector.

\noindent
{\bf Acknowledgments}
We thank W.N. Brandt, F. Halzen, 
D.P. Schneider, S. Razzaque, E. Waxman and B. Zhang 
for discussions. This research is supported in part by NSF AST 0307376,
the Monell foundation, and (for J.A-M) the Spanish ``Ministerio de 
Educaci\'on y Ciencia" through the ``Ram\'on y Cajal" program.

\end{document}